\documentstyle[preprint,prd,aps,psfig]{revtex}\tighten
\newcommand{\beq}{\begin{equation}}
\newcommand{\eeq}{\end{equation}}
\newcommand{\beqa}{\begin{eqnarray}}
\newcommand{\eeqa}{\end{eqnarray}}

\newcommand{\g}{ {\rm g} }
\newcommand{\cm}{ {\rm cm} }
\newcommand{\s}{ {\rm s} }

\newcommand{\erg}{ {\rm erg} }

\newcommand{\nick}{^{56}{\rm Ni}}
\newcommand{\co}{^{56}{\rm Co}}
\newcommand{\iron}{^{56}{\rm Fe}}
\newcommand{\msolar}{M_{\odot}}

\begin{document}
\widetext
\draft

\preprint{\tighten\vbox{\hbox{KUNS-1829}\hbox{RESCEU-4/03}\hbox{astro-ph/0306486}}}

\title{
Supernova Cosmology and the Fine Structure Constant
}

\author{Takeshi Chiba}
\address{
Department of Physics, Kyoto University, Kyoto 606-8502, Japan
}

\author{Kazunori Kohri}
\address{
Research Center for the Early Universe (RESCEU), University of Tokyo,
Tokyo 113-0033, Japan
}

\date{March 24, 2003}

\maketitle

\begin{abstract} 
We study the dependence of the peak luminosity of Type Ia supernovae 
on the fine structure constant $\alpha$. We find that
decreasing (increasing)  $\alpha$ enhances (reduces) the luminosity. 
Future experiments like SNAP could determine the variation of $\alpha$ to
a precision of $10^{-2}$.  
\end{abstract}

\section{Introduction}
Recently, the possibility of the variation of fundamental ``constants''
(especially the fine structure constant $\alpha$ and the gravitational
constant $G$) has attracted much attention.\cite{chiba}.  
Recent evidence that the value of $\alpha$ was smaller 
in the  past from observations of a number of absorption systems
in the spectra  of distant quasars has stimulated interest in this 
topic.\cite{Webb:2001mn,Webb:2002} {}From an anthropic point of view,
it is also intriguing to examine how changing physical
constants would affect the structure of the universe we observe.

The magnitude of the variation of $\alpha$ over time on a cosmological 
scale is constrained by several observations: those of the 
absorption/emission lines of distant objects, yielding information on 
the variation of fine-splitting ($0.5<z<4$), those of spectrum of cosmic 
microwave background anisotropies, yielding information on the variation 
 of the recombination process ($z=1000$), and those of Big Bang 
nucleosynthesis, yielding information on the variation of the masses of the 
proton and neutron ($z=10^{10}$). 
In this paper, we consider another method to constrain the cosmological
time variation of $\alpha$, observations of Type Ia supernovae. 
A Type Ia supernova is considered to be a good standard candle, because 
its peak luminosity correlates with the rate of decline of the magnitude. 
Observations of Type Ia supernovae have been used to constrain  
cosmological parameters.\cite{hzt,scp} The homogeneity of the peak 
luminosity is essentially due to the homogeneity of the progenitor mass, and 
this is primarily determined by the Chandrasekhar mass, which is 
proportional to $ G^{-3/2}$. 
The peak luminosity also depends on the diffusion time of photons,  
which depends on $\alpha$ through the opacity. A decrease in opacity reduces 
the diffusion time, allowing trapped radiation to escape more rapidly,
leading in turn to an increase in the luminosity.

\section{Type Ia supernovae and $\alpha$}

Now we consider the effect of changing $\alpha$ on the bolometric
absolute magnitude of the peak luminosity of Type Ia supernovae.  
We limit our study to the peak luminosity, because the physics behind the
relation between the peak luminosity and the rate of its decline is not
fully understood (see Ref.~\cite{pinto} for a recent
attempt). We write the energy deposition rate from the
$^{56}$Ni~$\to$~$^{56}$Co~$\to$~$^{56}$Fe decay chain inside the 
photosphere of Type Ia supernovae as
\begin{eqnarray}
    \label{eq:L_bol}
    F(t) = M_{56} \ q(t),
\end{eqnarray}
where $M_{56}$ is the mass of $^{56}$Ni in grams ($\sim 0.6\msolar \sim
1.2 \times 10^{33}\g$),~\cite{nomoto:1984,thielemann:1986} and $q(t)$
is given by~\cite{gehrels:1987}
\begin{eqnarray}
    \label{eq:q_t}
    q(t)= \left[S^{\gamma}_{\rm Ni} \ e^{-t/\tau_{\rm
      Ni}}+S^{\gamma}_{\rm Co} \left(e^{-t/\tau_{\rm Co}} -
        e^{-t/\tau_{\rm Ni}}\right)\right] f_{\rm dep}^{\gamma}(t) +
    S^{\beta}_{\rm Co} \left(e^{-t/\tau_{\rm Co}} - e^{-t/\tau_{\rm
      Ni}}\right),
\end{eqnarray}
with
\begin{eqnarray}
    \label{eq:Sgamma3}
    S^{\gamma}_{\rm Ni} &=& 4.03\times10^{10} \erg \ \s^{-1}
    \left(\tau_{\rm Ni}/8.51 \ {\rm days}\right)^{-1}, \nonumber \\
    S^{\gamma}_{\rm Co} &=& 6.78 \times 10^{9}\erg \ \s^{-1}
    \left(\tau_{\rm Co}/111.5 \ {\rm days}\right)^{-1}, \nonumber \\
    S^{\beta}_{\rm Co} &=& 0.232 \times 10^{9}\erg \ \s^{-1}
    \left(\tau_{\rm Co}/111.5 \ {\rm days}\right)^{-1}. \nonumber
\end{eqnarray}
In Ref.~\cite{table_isotopes:1996}, the most recent value of the mean
lifetime of $\nick$ ($\co$), which is determined by the weak interaction,
is stated to be $\tau_{\rm Ni} $= 8.51 days ($\tau_{\rm Co}$ = 111.5
days). For simplicity, in this paper we assume that the differences in energy 
between the exited states and the ground states of $\nick$,
$\co$ and $\iron$ are mainly determined by nuclear forces; the Coulomb
part in excitation energy is assumed to be small.  In general, this 
assumption is almost valid.  The $\gamma$-ray deposition function, $f_{\rm
  dep}^{\gamma}(t)$,  i.e., the fraction of $\gamma$-ray energy
deposited in supernova matter, is fitted by~\cite{colgate:1980}
\begin{eqnarray}
    \label{f_dep}
    f_{\rm dep}^{\gamma}(t) = G(\tau)\left(1 + 2G(\tau)\left[1-
        G(\tau)\right]\left[1-0.75G(\tau)\right] \right),
\end{eqnarray}
with
\begin{eqnarray}
    \label{G_tau}
    G(\tau) = \tau / \left(\tau + 1.6\right),
\end{eqnarray}
where $\tau = \tau(t)$ is the optical depth.

The peak luminosity of the optical light curve is essentially proportional to 
the value of $F(t)$ at a time $t_{p}$, when the expansion timescale is equal 
to the diffusion timescale:  $L_{\rm peak} \sim F(t=t_{p})$ with
$t_{p} \sim t_{\rm exp} \sim t_{\rm diff}$. Here the diffusion
timescale $t_{\rm diff}$ is given by
\begin{eqnarray}
    \label{t_diff}
    t_{\rm diff} = \kappa \rho R^{2} /c,
\end{eqnarray}
where $\kappa$ is the mean opacity ($\sim 0.1 \ \cm^{2} \  \g^{-1}$ in
Ref.~\cite{hoflich:1991}), $\rho$ is the matter density, and $R$ is
the radius. On the other hand, the expansion timescale $t_{\rm exp}$
is given approximately by
\begin{eqnarray}
    \label{t_expand}
    t_{\rm exp} = R / v,
\end{eqnarray}
where $v$ is the expansion velocity of the matter.  The total
mass of the progenitor is determined by the Chandrasekhar mass and is
given by $M = 4 R^{3} \rho/3\pi \simeq 3{\mu_e}^{-2}
G^{-3/2}m_p^{-2}\simeq 1.4\msolar$, where $\mu_e$ is the mean
molecular weight of electrons and $m_p$ is the proton
mass.\footnote{
Changing $\alpha$ causes a change in the nucleon mass through 
electromagnetic radiative corrections,\cite{gasser,tk} and hence a
change in the Chandrasekhar mass. However, the resulting change in the
luminosity is found to be smaller by four orders of 
magnitude than the value given in Eq.(\ref{eq:DeltaL_L}).
}
The total energy of the explosion is $E = M v^{2}/2 \sim 10^{51}$ erg, 
which is due to the difference between the binding energies of 
$\nick$ and C ($\sim$ 1 MeV times the number of nucleons).  Through the 
relation \beq t_{\rm diff} = {3\kappa M\over 4\pi c R} = {3\kappa
M\over 4\pi c V t_{\rm exp}}= {3\kappa M^{3/2}\over 4\sqrt{2}\pi c
E^{1/2}t_{\rm exp}}, \eeq we obtain
\begin{eqnarray}
    \label{t_p}
    t_{p} &\sim& 
    \left(\frac3{4\sqrt{2}\pi}(\kappa/c)\right)^{1/2}
    \left(\frac{M^{3}}{E}\right)^{1/4}, \\
    &\sim& 
    19~{\rm days} \left(\frac{\kappa}{0.1 \cm^{2} \
      \g^{-1}}\right)^{1/2}\left(\frac{M}{1.4\msolar}\right)^{3/4}
    \left(\frac{E}{10^{51} \erg}\right)^{-1/4}.
\end{eqnarray}
In this case we find that $L_{\rm peak} \sim 0.97 \times 10^{43} \erg
\ \s^{-1}$, and the velocity is given by $v \simeq 8.5 \times
10^{8} \ \cm \ \s^{-1} \left(E/10^{51} \ \erg\right)^{1/2}\left(M/1.4
  M_{\odot}\right)^{-1/2}$.

The mean opacity $\kappa$ should be proportional to $\alpha^{n}$ because
emitted photons scatter off ions through the electromagnetic
interaction ($n = 2$ for Thomson scattering). Because the density in the
supernova shell is very low ($\sim 10^{-13} \g \cm^{-3}$) after 10
days, the opacity may be almost entirely due to electron scattering ($n =
2$)\cite{cs}. However, for generality we include the $n$
dependence.~\footnote{
A photon in the expanding shell suffers a continuous Doppler shift of
frequency with respect to the rest frame of the material. Those
photons which are redshifted to the frequency of a sufficiently strong
line will be absorbed by the corresponding bound-bound transition.
This effect would effectively increase the opacity~\cite{karp:1977,muller:1991}. Such an enhancement factor
has a complicated form as a function of the bound-bound transition
opacity and is non-linear in $\alpha$~\cite{karp:1977}.  Near the peak
luminosity, however, the effective value of $n$ modified by the
enhancement factor is the same order of magnitude, and therefore this 
effect should not affect the following analysis significantly.
} 
Then, the uncertainty in $t_{p}$ is related with the change of $\alpha$
as
\begin{eqnarray}
    \label{eq:dt_p}
    \frac{\Delta t_{p}}{t_{p}} = \frac12 \frac{\Delta \kappa}{\kappa}
    = \frac{n}{2}\frac{\Delta \alpha}{\alpha}.
\end{eqnarray}
{}From Eq.~(\ref{eq:q_t}), we see that the uncertainty in $q(t_{p})$ caused
by the variation of $ t_{p}$ or $\alpha$ can be expressed by
\begin{eqnarray}
    \label{eq:deltaq_t}
    \Delta q(t_{p}) &=& - \frac{\Delta t_{p}}{\tau_{\rm
    Ni}}\left[S^{\gamma}_{\rm Ni}f_{\rm
      dep}^{\gamma}(t_{p})e^{-t_{p}/\tau_{\rm
      Ni}}+\left(S^{\gamma}_{\rm Co}f_{\rm
        dep}^{\gamma}(t_{p})+S^{\beta}_{\rm Co} \right)\left(
        \frac{\tau_{\rm Ni}}{\tau_{\rm Co}}e^{-t_{p}/\tau_{\rm Co}} -
        e^{-t_{p}/\tau_{\rm Ni}}\right) \right] \nonumber \\
    &&+ \Delta f_{\rm dep}^{\gamma}(t_{p}) \left[S^{\gamma}_{\rm
      Ni}e^{-t_{p}/\tau_{\rm Ni}}+S^{\gamma}_{\rm Co}\left(
        \frac{\tau_{\rm Ni}}{\tau_{\rm Co}}e^{-t_{p}/\tau_{\rm Co}} -
        e^{-t_{p}/\tau_{\rm Ni}}\right) \right],
\end{eqnarray}
where
\begin{eqnarray}
    \label{eq:Delta_f_dep}
    \Delta f_{\rm dep}^{\gamma}(t_{p}) = \Delta G(\tau_{p})\left[1 +
      4G(\tau_{p}) - 10.5G(\tau_{p})^{2}+6.0 G(\tau_{p})^{3} \right],
\end{eqnarray}
with $\tau_{p} \equiv \tau(t=t_{p})$, and
\begin{eqnarray}
    \label{eq:Delta_G}
    \Delta G(\tau_{p}) = \frac{1.6}{\left(1.6 + \tau_{p}\right)^{2}}
    \Delta \tau_{p}.
\end{eqnarray}
In our simple treatment, the optical depth ($\tau = \kappa \rho R$)
is assumed to be proportional to $\kappa \ t^{-2} \ v^{-2} \ M$. If we 
adopt the value
of $\tau$ in Ref.~\cite{colgate:1980} ($\tau \simeq$ 1.0 at $t$ = 20
days for $v= 1.5 \times 10^{9} \ \cm \ \s^{-1}$), we obtain
\begin{eqnarray}
    \label{eq:tau_p}
    \tau_{p} \sim 3.6 \left(\frac{E}{10^{51} \erg}\right)^{-1/2}
    \left(\frac{M}{1.4\msolar}\right)^{1/2}.
\end{eqnarray}
{}From Eq.~(\ref{eq:tau_p}), we see that $\tau_{p}$ does not depend on
$\kappa$, and the second term in Eq.~(\ref{eq:deltaq_t}) disappears.
We note that in this case $f_{\rm dep}^{\gamma}(t_{p}) \sim
0.83$.

Using Eqs.~(\ref{eq:L_bol}),~(\ref{eq:dt_p}) and (\ref{eq:deltaq_t}),
we obtain the relation
\begin{eqnarray}
    \label{eq:DeltaL_L}
    \frac{\Delta L_{\rm peak}}{L_{\rm peak}} = \frac{\Delta
    q(t_{p})}{q(t_{p})} = -  a \frac{n}{2} \frac{\Delta \alpha}{\alpha},
\end{eqnarray}
where $a \sim 0.94$. Thus, we obtain the following relationship
between $\Delta \alpha$ and the change in the absolute magnitude $\Delta
{\cal M}$ at the peak luminosity:
\beq
    \label{eq:DL_L}
  {\Delta\alpha\over \alpha}={4 \ln 10\over 5 a n}\Delta {\cal M}
  =0.98\left({0.94\over a}\right) \left({2\over 
  n}\right)\Delta {\cal M}.
\eeq
This is the main result of this paper. 
It can be understood as follows. Decreasing $\alpha$
causes the opacity to decrease, which allows photons to escape more 
rapidly, thereby leading to an increase in the luminosity. 
Thus a smaller (larger) value of $\alpha$ would make supernovae 
brighter (fainter). 

Let us now estimate the severity of the constraint derived from observations 
of Type Ia supernovae. 
Future experiments like SNAP(SuperNova/Acceleration
Probe)\footnote{http://snap.lbl.gov} ae planned to observe 
thousands of supernovae and should be able to reduce systematic errors to a 
magnitude of $0.02$ mag, 
which corresponds to $\Delta\alpha/\alpha < 2\times 10^{-2}$. 
This is  larger than the current limit for 
$0.16<z<0.80$, $\Delta\alpha/\alpha =(-2\pm
1.2)\times 10^{-4}$,\cite{ba} although the limit obtained from supernovae 
could apply to higher redshifts and depends on the nature of the time 
evolution of $\alpha$.  

The sensitivity of this ``supernova method'' to the variation of $\alpha$ 
might be on the same order as that of CMB \cite{hannestad99}. 
(For the case of  varying $G$ see \cite{gazta} and \cite{ncs}). 
In any case, it is important to determine the possible limits of the 
variation of fundamental constants by various means. As one such 
effort, we have proposed a method to determine the limit of the time 
variation of $\alpha$ through examination of 
the variation of the peak luminosity of supernovae. 
Detailed analysis will be published elsewhere.

\section*{Acknowledgements}
K.K. wishes to thank T.~Shigeyama and N.~Itagaki for useful
discussions. T.C. was supported in part by a Grant-in-Aid for Scientific 
Research (No.13740154) from the Japan Society for the Promotion of
Science and by a Grant-in-Aid for Scientific Research on Priority Areas 
(No.14047212) from the Ministry of Education, Science, Sports and 
Culture, Japan.


\end{document}